\documentclass[dvips]{acta}
\usepackage{supertabular,lscape,epsfig}
\usepackage{amssymb}
\usepackage{amsmath}
\SetPages{1}{12}
\SetVol{59}{2009}

\begin{document}

\begin{Titlepage}

\Title { Z Cha and its Superhumps }

\Author {J.~~S m a k}
{N. Copernicus Astronomical Center, Polish Academy of Sciences,\\
Bartycka 18, 00-716 Warsaw, Poland\\
e-mail: jis@camk.edu.pl }

\Received{  }

\end{Titlepage}

\Abstract{ 
The superhump eclipse light curves are re-determined for five eclipses of 
Z Cha (E54036, E54037, E67877, E74693, E74694) observed by Warner and 
O'Donoghue (1988) during its superoutbursts. Qualitatively they are similar 
to those obtained by O'Donoghue (1990), showing two local minima at 
$\phi\sim-0.05$ and 0.04. 
Arguments are then presented which imply that the first minimum is not due 
to an occultation but is produced by absorption effects in the overflowing parts 
of the stream. 
The location of the superhump light source (SLS) determined from the analysis 
of the second minimum coincides with the trajectory of the overflowing parts 
of the stream. 

The light curve of the sixth eclipse (E77878) could be simply decomposed into 
its disk and superhump components. The location of SLS, obtained from the 
analysis of the SLS eclipse light curve, coincides in this case with the position 
of the standard hot spot. 

This implies that superhumps are due to modulated mass transfer rate resulting 
in periodically enhanced dissipation of the kinetic energy of the stream. }
{accretion, accretion disks -- binaries: cataclysmic
variables, stars: dwarf novae, stars: individual: Z Cha }

\section {Introduction }

This is the third paper in a series devoted to the analysis of light curves 
of Z Cha observed during its superoutbursts by Warner and O'Donoghue (1988). 
In the first paper (Smak 2007) the light curves covering eclipses located 
away from superhumps were decomposed into their disk eclipse and hot 
spot eclipse components. 
In the second paper (Smak 2008) the accretion rates, determined from disk 
eclipse analysis, were found to be practically identical with the mass 
transfer rates determined from spot luminosities, what implies that 
superoutbursts are due to a major enhancement in the mass transfer rate. 
In the present paper we analyze eclipses at beat phases near 
$\phi_b \sim 0$, i.e. those which involve the occultation of the superhump 
light source. 

The first such analysis was made by Warner and O'Donoghue (1988) who applied 
the Maximum Entropy Method (MEM) to two eclipses involving superhumps: 
E54037 and E77878. In the first case the superhump light source (SLS) was 
identified by them with a peak in the resulting "MEM image", located near 
the edge of the disk and facing the secondary component (i.e. not far from 
the location of the standard hot spot). In the case of E77878 they found 
the location of SLS to be "{\it consistent with a position on the edge 
of the disk beginning at the quiescent bright spot and continuing downstream}" 
or "{\it along the continuation of the stream into the inner disc}". 
In spite of that, however, Warner and O'Donoghue concluded that those  
results "{\it eliminate locations in the vicinity of the quiescent bright spot 
or along the mass-stream}"... 

Two years later O'Donoghue (1990), using also the MEM technique, made a more 
detailed analysis of four superhump eclipses. 
His "MEM images" for E54036, E54037, and E67877 showed that the superhump light 
source (SLS) consists of three distinct areas: 
area 1 -- on the trailing lune of the disk, area 2 -- on its leading lune, 
and area 3 -- near the edge of the disk, facing the secondary component. 
This was interpreted by O'Donoghue (and emphasized in the title of his paper) 
as the observational evidence for the tidal origin of superhumps. 
This interpretation, however, encounters one serious problem.  
Contrary to the statements by O'Donoghue that "{\it the superhump light source 
is located on the rim of the disk}" and that this "{\it strongly suggests that 
tidal stresses are responsible}", one can easily see from his "MEM images" 
that the centers of area 1 and area 2 are located at $r/r_d\approx 0.5-0.6$ 
(where $r_d=r_{tid}=0.9r_{Roche}$), i.e. roughly half-way between disk edge and 
its center. In addition, there was the case of E77878 showing -- as before -- 
that SLS consists of only one area located "{\it very close to the position 
of the bright spot at quiescence}". 

Those discrepancies suggested that another, independent analysis of the problem 
should be undertaken. 
In Section 2 we substantially modify the method used earlier by O'Donoghue to 
obtain the superhump eclipse light curves and present the resulting new light 
curves for five eclipses (E54036, E54037, E67877, E74693, and E74694). 
They are qualitatively similar to those obtained by O'Donoghue, showing two 
local minima at $\phi\sim -0.05$ and $\phi\sim 0.04$. 
Evidence is then presented (Sections 3 and 4) which implies that the first 
minimum is produced by absorption effects in the overflowing parts of the 
stream. From the analysis of the second minimum (Section 5) we find that 
the location of SLS coincides with the trajectory of the overflowing parts 
of the stream. In Section 6, devoted to the sixth eclipse E77878, we simply 
decompose its observed light curve into the disk and superhump components  
and find that in this case the location of SLS coincides with the position 
of the standard hot spot. Using this evidence we conclude (Section 7) that 
(1) superhumps are due to periodically enhanced dissipation of the kinetic energy 
of the stream resulting from strongly modulated mass transfer rate and  
(2) substantial stream overflow occurs around superhump maximum making the superhump 
light source similar to "peculiar" spots observed at intermediate beat phases.

\section {The Light Curves of Superhumps }

O'Donoghue (1990) analyzed five eclipses (E54036, E54037, E67877, E74693 and E77878) 
involving superhumps using the following simple method. 
The superhump eclipse light curve was determined as 

\beq
\ell_{sh}~=~\ell_{obs}({\rm d})~-~\ell_{obs}({\rm d}-1)~,
\eeq

\noindent
where $\ell_{obs}({\rm d})$ is the eclipse light curve under analysis 
and $\ell_{obs}({\rm d}-1)$ -- the eclipse light curve observed on the 
previous night, when the eclipse was located away from superhumps. 
The out-of-eclipse superhump light curve $\ell_{sh}^{~\circ}$ was assumed 
to be identical with that observed also on the previous night. 
The resulting superhump eclipse light curves (Figs.1-7 in 
O'Donoghue 1990) showed two local minima: one near $\phi\sim -0.05$, 
and another at $\phi\sim 0.05$, and a local maximum between them near 
$\phi\sim 0.0$. 

The method used by O'Donoghue suffered from his arbitrary assumption 
that the light curves observed on the previous night are representative 
for the situation under analysis. This is not the case. 
First of all, the observed depth of eclipse increases with time 
(see Fig.6 in Warner and O'Donoghue 1988). Secondly, the eclipse 
light curve observed on the previous night (away from superhump) 
consists of the disk and spot components, the contribution from the 
spot depending strongly on the superhump phase (cf. Smak 2007). 
Thirdly, the superhump amplitude is also not constant: it decreases 
with time (see Fig.5 in Warner and O'Donoghue 1988). 

In the present analysis we modify the method used by O'Donoghue in 
several ways. To begin with, we determine the superhump eclipse 
light curve as 

\beq
\ell_{sh}~=~\ell_{obs}~-~\ell_d~,
\eeq

\noindent
where $\ell_d$ is the pure disk eclipse curve. As shown earlier 
(Smak 2008, Fig.1), the depth of disk eclipse  $\Delta \ell_d$ increases 
with time. Rewriting Eq.(1) from that paper we have 

\beq
\Delta \ell_d~=~0.494~+~0.050~\Delta t~,
\eeq

\noindent
where $\Delta t$ is the time (in days) since the beginning of 
superoutburst.
To check whether the shapes of disk eclipse light curves differ also in some 
other way, we reduce all disk light curves obtained earlier from eclipses 
observed at beat phases $0.4<\phi_b<0.6$ (Smak 2007) to an arbitrarily adopted central intensity $\ell_{\circ}=0.50$. 
Results, presented in Fig.1, show that the shapes of eclipses normalized 
in such a way are practically identical. 
This mean reduced disk light curve together with $\Delta \ell_d$ from Eq.(3) 
can then be used to calculate the disk light curve applicable via Eq.(2) 
to the situation considered.

\begin{figure}[htb]
\epsfysize=6.0cm 
\hspace{2.0cm}
\epsfbox{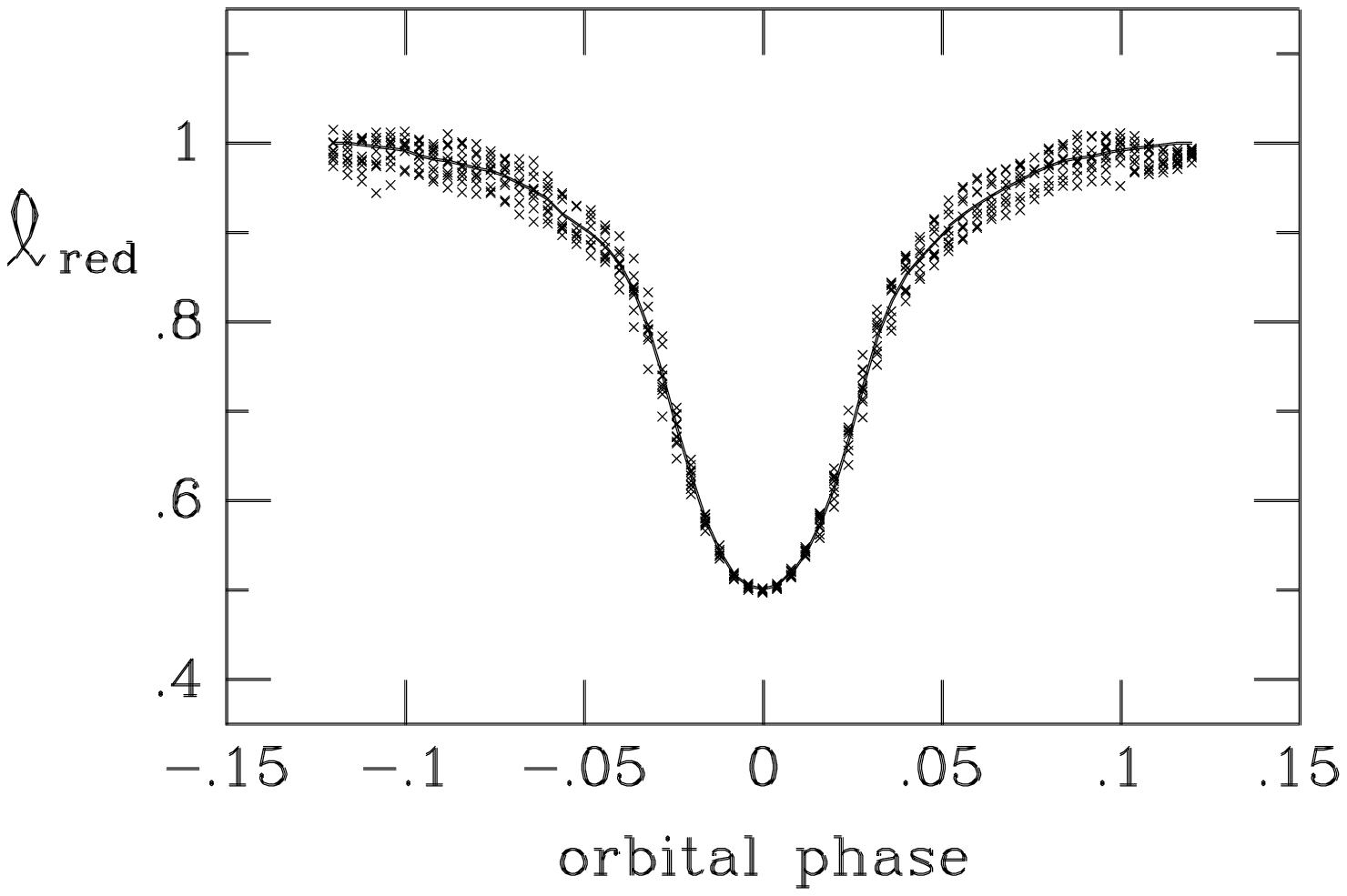} 
\vskip 5truemm
\FigCap { Disk eclipse light curves reduced to central intensity
$\ell_{\circ}=0.5$. Solid line is the mean reduced curve. }
\end{figure}

To obtain the normalized superhump light curve  

\beq
\ell_{sh}^{~n}~=~\ell_{sh}/\ell_{sh}^{~\circ}
\eeq

\noindent
we need the out-of-eclipse superhump light curve $\ell_{sh}^{~\circ}$. 
This step inroduces some uncertainty. In addition to the already mentioned dependence of the superhump amplitude on $\Delta t$, their shapes vary considerably from one superhump to another showing, in particular, 
large scale rapid flickering. 
After analyzing superhump light curves from Warner and O'Donoghue (1988) 
it was found (in agreement with a similar conclusion by O'Donoghue 1990) 
that their shapes can be -- approximately -- represented by 

\beq
\ell_{sh}^{~\circ}(\phi)~=~A_{sh}~\exp ~[-a(\phi-\phi_{max})^2]~,
\eeq

\noindent
or, to account for their frequent asymmetry, by

\beq
\ell_{sh}^{~\circ}(\phi)~=~
     A_{sh}~\exp ~[-a(\phi-\phi_{max})^2~-~b(\phi-\phi_{max})^3]~,
\eeq

\noindent
where $A_{sh}$ is the superhump amplitude and $\phi_{max}$ -- 
the phase of its maximum. 
Depending on the situation (see below) one of these two formulae   
was fitted to the out-of-eclipse parts of the observed superhump 
light curves. 

In this Section we present results for five eclipses, namely: 
E54036, E54037, E67877, E74693 and E74694 (not analyzed by O'Donoghue). 
In the first two cases, with superhump maximum occuring within 
the eclipse, we use Eq.(6) with $A_{sh}=0.35$ (from Fig.5 of 
Warner and O'Donoghue 1988), $\phi_{max}$ from the superhump ephemeris, 
and determine the two remaining unknown parameters: $a$ and $b$. 
In the three other cases, with superhump maximum occuring just before 
the eclipse, we use Eq.(5) and determine all three parameters: 
$A_{sh}$, $\phi_{max}$ and $a$. 

Results are presented in Figs.2-4. As can be seen our superhump 
light curves are qualitatively similar to those of O'Donoghue:  
they show two local minima at $\phi\sim -0.05$ and $\phi\sim 0.04$, 
and a local maximum near $\phi\sim -0.01$.

\begin{figure}[htb]
\epsfysize=11.0cm 
\hspace{2.5cm}
\epsfbox{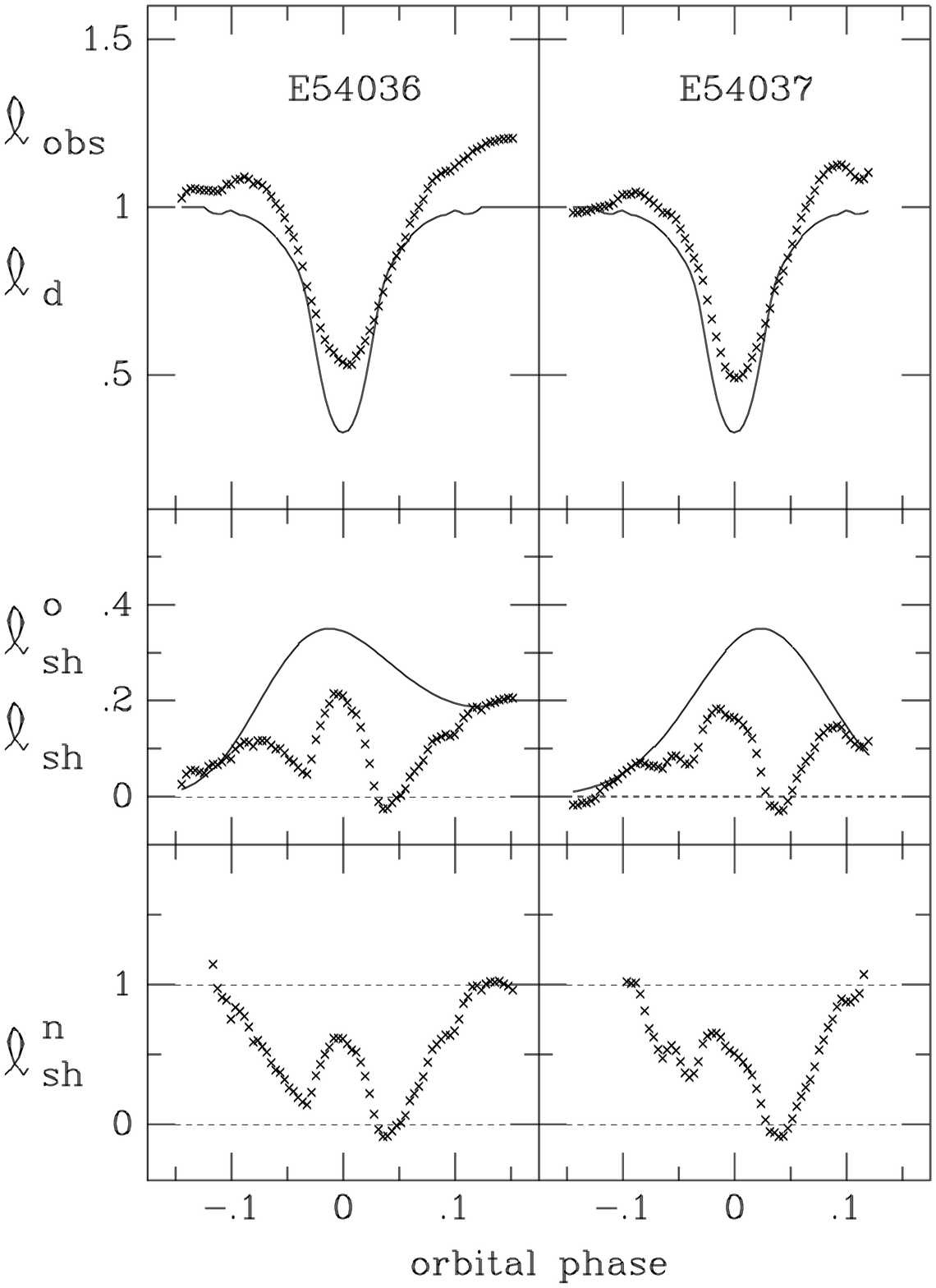} 
\vskip 5truemm
\FigCap { Light curve analysis for eclipses E54036 and E54037. 
Shown with crosses are the observed light curves ({\it top}),  
the superhump light curves ({\it middle}), and the normalized 
superhump light curves ({\it bottom}). 
Shown with solid lines are the disk light curves ({\it top}) 
and the fitted out-of-eclipse superhump light curves ({\it middle}). }
\end{figure}

\begin{figure}[htb]
\epsfysize=11.0cm 
\hspace{2.5cm}
\epsfbox{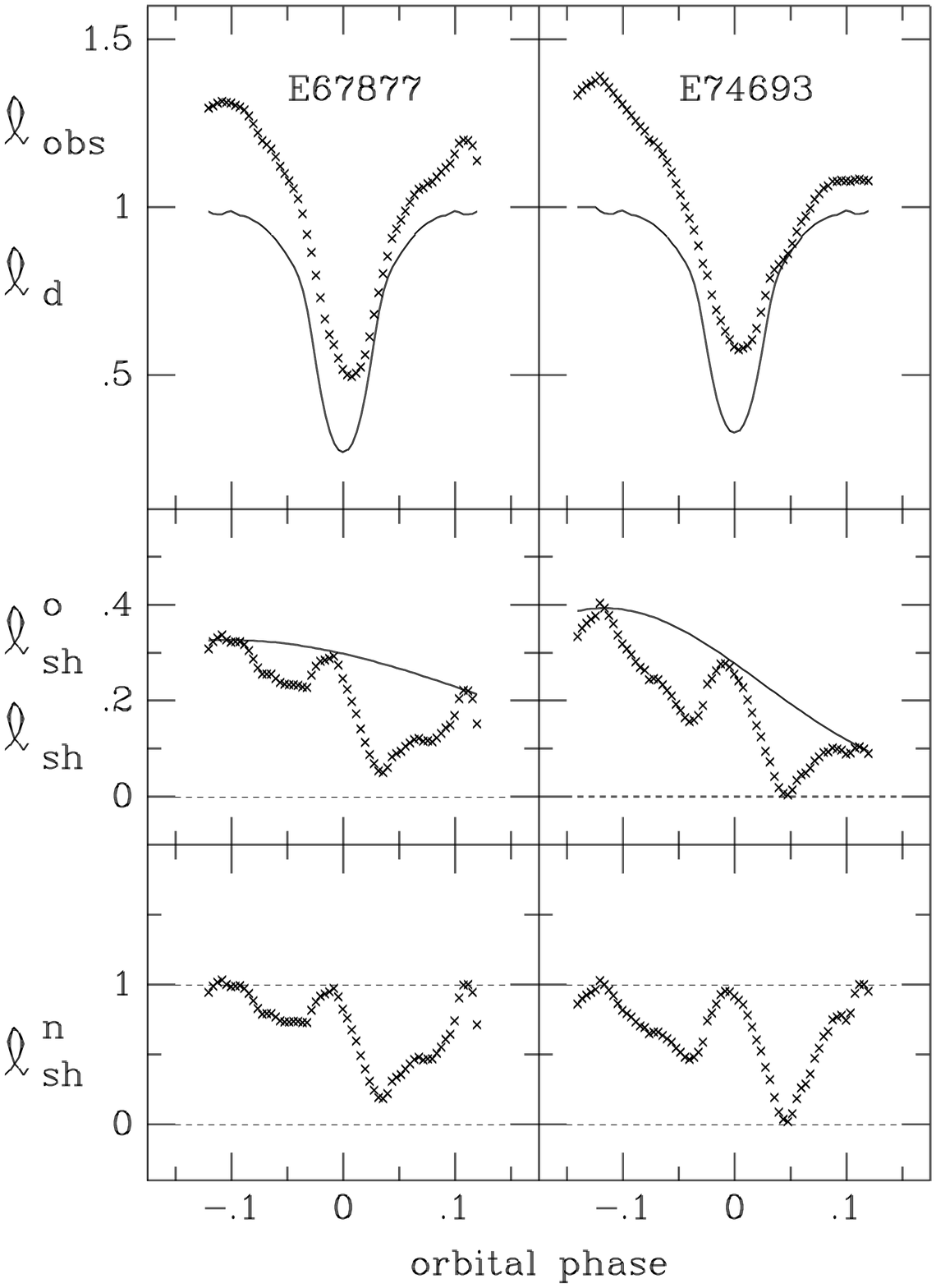} 
\vskip 5truemm
\FigCap { Light curve analysis for eclipses E67877 and E74693 (see caption to Fig.2). }
\end{figure}

\begin{figure}[htb]
\epsfysize=11.0cm 
\hspace{2.5cm}
\epsfbox{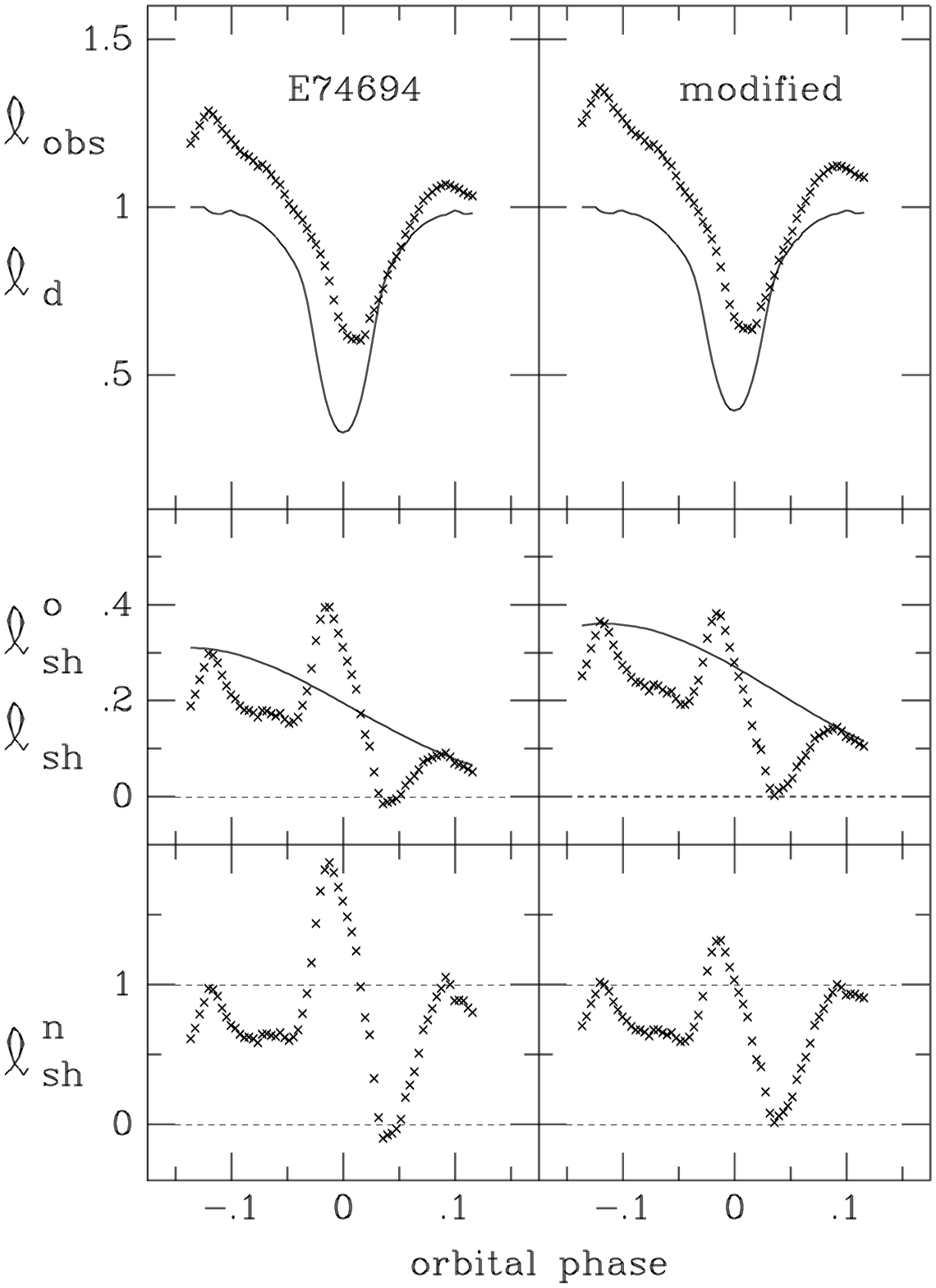} 
\vskip 5truemm
\FigCap { Light curve analysis for the eclipse E74694 (see caption to Fig.2).   
The left panel shows the standard analysis while the right panel -- 
its modification, as discussed in the text. }
\end{figure}

The case of E74694 (not analyzed by O'Donoghue) requires special comments. 
The locations and depths of the two local minima are similar to other 
cases but the local maximum at $\phi \sim -0.01$ reaches an unacceptable 
value of $\ell_{sh}^{~n}({\rm max})\approx 1.9$.  
In an attempt to explain its origin we repeated our analysis using 
slightly modified input parameters. 
The right panel of Fig.4 shows an example of such a modification 
in which the observed light curve was corrected by adopting "level 1" 
at $\ell =0.95$ and the disk light curve was made shallower by 10 percent 
(both these changes being acceptable within existing uncertainties). 
As one can see, those modifications resulted in a much lower height 
of the local maximum (the shape of the two minima remaining roughly 
the same). In view of the arbitrary nature of such modifications, 
however, no attempt was made to continue such experiments until 
getting $\ell_{sh}^{~n}({\rm max})=1$. 

Turning to a more detailed comparison of our superhump light curves with 
those obtained by O'Donoghue, we note significant difference with respect 
to the second minimum (at $\phi \sim 0.04$) which in our case is much 
deeper: the four light curves of O'Donoghue (excluding E77878) give 
the mean depth of $<\Delta \ell> \approx 0.8$, while our five light curves 
give $<\Delta \ell> \approx 1.0$. There are also other differences. 
In the case of E67877 the local maximum around $\phi \sim -0.01$ obtained 
by O'Donoghue reached only $\ell_{sh}^{~n}\approx 0.7$, while in our case $\ell_{sh}^{~n}\approx 1.0$. 
In the case of eclipse E74693 O'Donoghue encountered problem similar to our problem 
with E74694: his local maximum around $\phi \sim -0.01$  exceeded "level 1". 
Our analysis of this eclipse removed this problem giving $\ell_{sh}^{~n}\approx 1.0$.

\section {Are We Dealing with Pure Eclipses? }

In the standard analysis of eclipse light curves they are used to determine 
the surface brightness distribution over the surface of the disk. 
O'Donoghue (1990) did this using the Maximum Entropy Method (MEM). 
From the very nature of the MEM technique one could expect the model light 
curves calculated from the "best MEM image" to provide perfect fit to the 
observed eclipse light curves. 
However, as can be seen from Figs.1-5 of O'Donoghue's, all his model 
light curves show minima which are too shallow by as much as 
10-20 percent. Obviously then something must have been wrong... 

In our analysis we used a much simpler and more straightforward method: 
The disk was divided into $6\times 6=36$ square areas and their surface 
brightnesses ($x_i,~i=1,36$) were determined by a formal least-squares 
fit to the observed light curve. 
In all cases analyzed the results were dramatically disappointing: 
roughly one-third of the resulting values of $x_i$ (particularly 
those on the leading lune of the disk) turned out to be negative 
or strongly negative! 
Again, obviously, something must have been wrong...

To identify the nature of those problems let us take a closer look 
at the light curves. As mentioned above, the mean depth of the second 
minimum at $\phi \approx 0.04$ is $<\Delta \ell> \approx 1.0$ what means 
that the superhump light source (SLS) is fully eclipsed at that phase. 
The relatively narrow shape of this minimum  and its central phase imply 
that SLS must be rather small and located somewhere on the trailing lune 
of the disk. If so, the first minimum, at $\phi\approx -0.05$, cannot 
be due to another occultation of SLS at that phase... 

Our arguments become even more straightforward in the case of 
eclipses E67877, E74693, and E74694. At $\phi \approx -0.01$ their light 
curves (Figs.3 and 4) show $\ell_{sh}^{~n} \approx 1$. This could imply 
that the two minima represent two separate eclipses of two parts of SLS. 
If so, their normalized depths should obey the obvious condition:   
$\Delta \ell_1+\Delta \ell_2 \leq 1$. 
In fact, however, in the case of those three eclipses we have, respectively,  
$\Delta \ell_1+\Delta \ell_2 \approx 1.1$, 1.5, and 1.4. 
This means that they cannot be due only to an occultation of SLS by the 
secondary component. In other words -- that one of the two minima (or both) 
must be partly due to some other effect(s). 
This explains why previous attempts to analyze those light curves 
by treating them as pure eclipses encountered problems discussed above.

\section { The Case for Absorption Effects }

Listed below are facts and arguments which consistently 
suggest that problems described above have their source in absorption 
effects. Specifically -- that the first minimum is caused 
by absorption in the overflowing parts of the stream. 

(1) The evidence for a substantial stream overflow in Z Cha during its 
superoutbursts came from the analysis of "peculiar" spot eclipses 
observed at intermediate beat phases away from $\phi_b \sim 0.5$ 
(Smak 2007). 
The spot distances obtained from those eclipses turned out to be 
smaller than the radius of the disk, indicating that such "peculiar" 
spots are formed partly in the overflowing parts of the stream. 

The stream overflow was originally expected to occur only when 
the disk is geometrically thin, i.e. mainly in quiescent dwarf novae 
(cf. Hessman 1999 and references therein). In addidtion to the 
observational evidence quoted above, however, there is also theoretical evidence (Kunze et al. 2001) suggesting that it is a much more common phenomenon. 

(2) Fig.5-{\it left} shows the view of the system at $\phi=-0.05$.  
It can immediately be seen that around that phase the overflowing 
portions of the stream pass between the observer and the center 
of the disk and therefore are likely to absorb part of the flux coming 
from its  central (brightest) parts. 
On the other hand, Fig.5-{\it right} shows that no major effects of 
this type should be present around the second minimum at $\phi=0.04$ 
simply because the stream and the adjacent portions of the disk 
are fully eclipsed at that phase.

\begin{figure}[htb]
\epsfysize=2.5cm 
\hspace{1.0cm}
\epsfbox{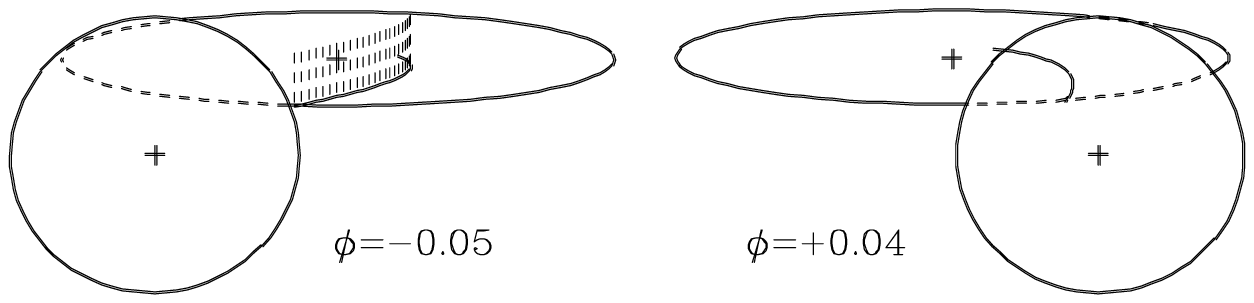} 
\vskip 5truemm
\FigCap { View of Z Cha at $\phi=-0.05$ ({\it left}) and $\phi=0.04$ ({\it right}). 
Solid lines show the disk, the secondary, and the stream trajectory 
in the orbital plane ($z=0$). 
Short dashed lines in the left picture show the vertical extent 
($z=0.1$, assumed) of the overflowing parts of the stream. }
\end{figure}

The effective optical depth of the absorbing material can be estimated  
from: 

\beq
\tau~=~\ln~\left( {{\xi~\ell_d~-~\Delta\ell_{sh}~\ell_{sh}^{~\circ}}
       \over {\xi~\ell_d}} \right)~, 
\eeq 

\noindent
where $\Delta \ell_{sh}$ is the depth of the first minimum at 
$\phi=-0.05$, $\ell_{sh}^{~\circ}$ -- the out-of-eclipse luminosity  
(in intensity units normalized to "level 1"), $\ell_d$ -- 
the luminosity of the disk at that phase, and $\xi$ -- its effective 
fraction affected by absorption.  
Using values of the relevant parameters taken from Figs.2 and 3 
and assuming $\xi=0.5$ we get a range of $\tau \approx 0.2-1.6$  
which shows that relatively litle material is needed to produce 
the observed effects. 

(3) The idea of absorption effects being produced in the overflowing 
parts of the stream is not new. The first evidence for such effects 
came from the analysis of "peculiar" spot eclipses (Smak 2007) 
where it was found that spot distances obtained from ingress $r_i$ 
are systematically larger than those obtained from egress $r_e$. 
This was interpreted as being due to selfabsorption in the overflowing 
parts of the stream: during ingress, when the stream trajectory is 
nearly parallel to the line of sight, the effective light center of 
the "peculiar" spot is observed closer to the disk edge. 

Two additional effects supporting this interpretation can be predicted. 
To do so let us recall some details of the method which was used 
to decompose the observed light curves into their disk and spot 
components (Smak 1994, 2007). 
In step 2, involving crucial assumption that the disk eclipse light 
curve is symmetric around phase zero, the spot light curve is 
determined in the phase interval $[-\phi_3, -\phi_2]$. 
Worth noting is that the values of $\ell_s$ for $\phi<\phi_1$, 
representing its uneclipsed portion, are essential for determining 
the spot amplitude $A_s$. 
Let us now consider the situation when absorption effects of the type 
discussed above are present, affecting the shape of the disk eclipse 
light curve at negative phases around $\phi \sim -0.05$. 
Since we now have $\ell_d(\phi<0)<\ell_d(\phi>0$ the resulting 
values of $\ell_s(\phi<0)$ will come out lower by 
$\Delta \ell_s=\ell_d(\phi>0)-\ell_d(\phi<0)$, causing the resulting 
spot amplitude $A_s$ to be also lower. 
Turning to the observational evidence we recall that spot amplitudes 
determined from "peculiar" spot eclipses (Smak 2007, Fig.5) are indeed 
systematically lower. 
  
Another effect can be predicted for the part of the spot light curve 
around $\phi=0$, representing its total eclipse, where we normally have $\ell_s(\phi)\equiv 0$. In the case of absorption effects, however, 
when $\ell_d(\phi<0)<\ell_d(\phi>0)$ we expect $d\ell_s/d\phi>0$. 
Turning to the observational evidence and using the "standard" spot 
light curves we get $d\ell_s/d\phi=-0.08\pm 0.06$. 
In the case of "peculiar" spot light curves, however, we obtain $d\ell_s/d\phi=+0.67\pm 0.08$. 
(Worth adding is that this difference could be seen directly from 
Fig.1 in Smak (2007) showing three examples of "standard" spots 
and another three examples of "peculiar" spots).

\section { The Location of the Superhump Light Source }

As already mentioned earlier, no major absorption effects are to be 
expected in the case of the second minimum around $\phi \approx 0.04$. 
Therefore it can be treated as being mainly due to an occultation 
of the superhump light source (SLS). 
Taking into account, however, that some absorption effects may still 
contribute to its shape we do not attempt to analyze it in any formal way. 
Instead we limit ourselves to a qualitative analysis.

\begin{figure}[htb]
\epsfysize=4.0cm 
\hspace{4.5cm}
\epsfbox{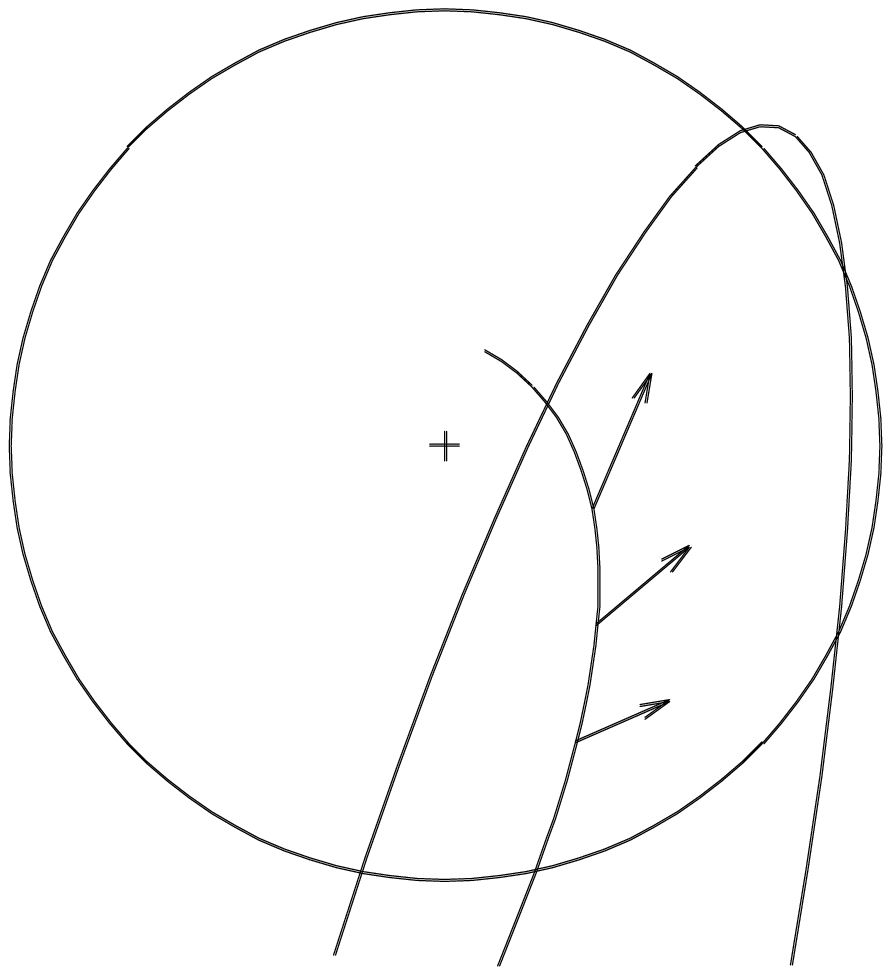} 
\vskip 5truemm
\FigCap { Geometrical constraints on the location of SLS. 
Shown are: the disk and the limb of the secondary at $\phi=0.04$ 
in projection on the orbital plane (broken lines). 
Solid line is the stream trajectory, while arrows represent the rotational 
velocity vectors of the disk. See text for details. }
\end{figure}

The geometry of eclipse at the central phase of the second minimum 
($\phi=0.04$) presented in Fig.6 shows that the eclipsed area includes the stream 
and the adjacent parts of the disk. At first sight the stream may appear not to 
be the best candidate since its trajectory is located asymmetrically with respect 
to the central line of the projected limb of the secondary. 
We should recall, however, that this trajectory was calculated without 
taking into account any interactions with the surface elements of the disk. 
It is obvious that due to those interactions the stream must be deflected 
in the direction of disk's rotation, as shown by arrows in Fig.6. 
Taking this into account we can conclude that the location of SLS actually 
coincides with the overflowing parts of the stream.

\section { The Case of E77878 }

This eclipse was described by Warner and O'Donoghue as "{\it anomalous}" or 
"{\it extremely peculiar}". It appears that the only reason for such a classification was that results obtained from this eclipse (quoted in the Introduction) did not support their main conclusion about tidal origin of superhumps. 

E77878 is, in fact, simpler and cleaner than other eclipses. 
To begin with, the superhump eclipse light curve (Fig.7 in O'Donoghue 1990) 
shows no trace of the first minimum around $\phi \sim -0.05$. 
This implies that the absorption efects (responsible for the first minimum 
observed in other cases) were absent. 
If so, it appeared reasonable to try to decompose the observed light 
curve into its disk and superhump component using the same simple 
method which was used earlier in the case of hot spots (Smak 2007). 
Results are shown in Fig.7. Filled squares represent points obtained 
in Steps 1 and 2 of the decomposition procedure (see Smak 1994). 
They are used to determine (using Eq.5) the "out-of-eclipse" superhump light 
curve (solid line), which is then used in Step 3 to determine the remaining 
points (open squares). 
The resulting light curve closely resembles those of hot spots: 
The eclipse is total, its ingress and egress are well defined, 
and the only obvious difference is that the luminosity of the 
superhump varies with time.

\begin{figure}[htb]
\epsfysize=6.0cm 
\hspace{1.5cm}
\epsfbox{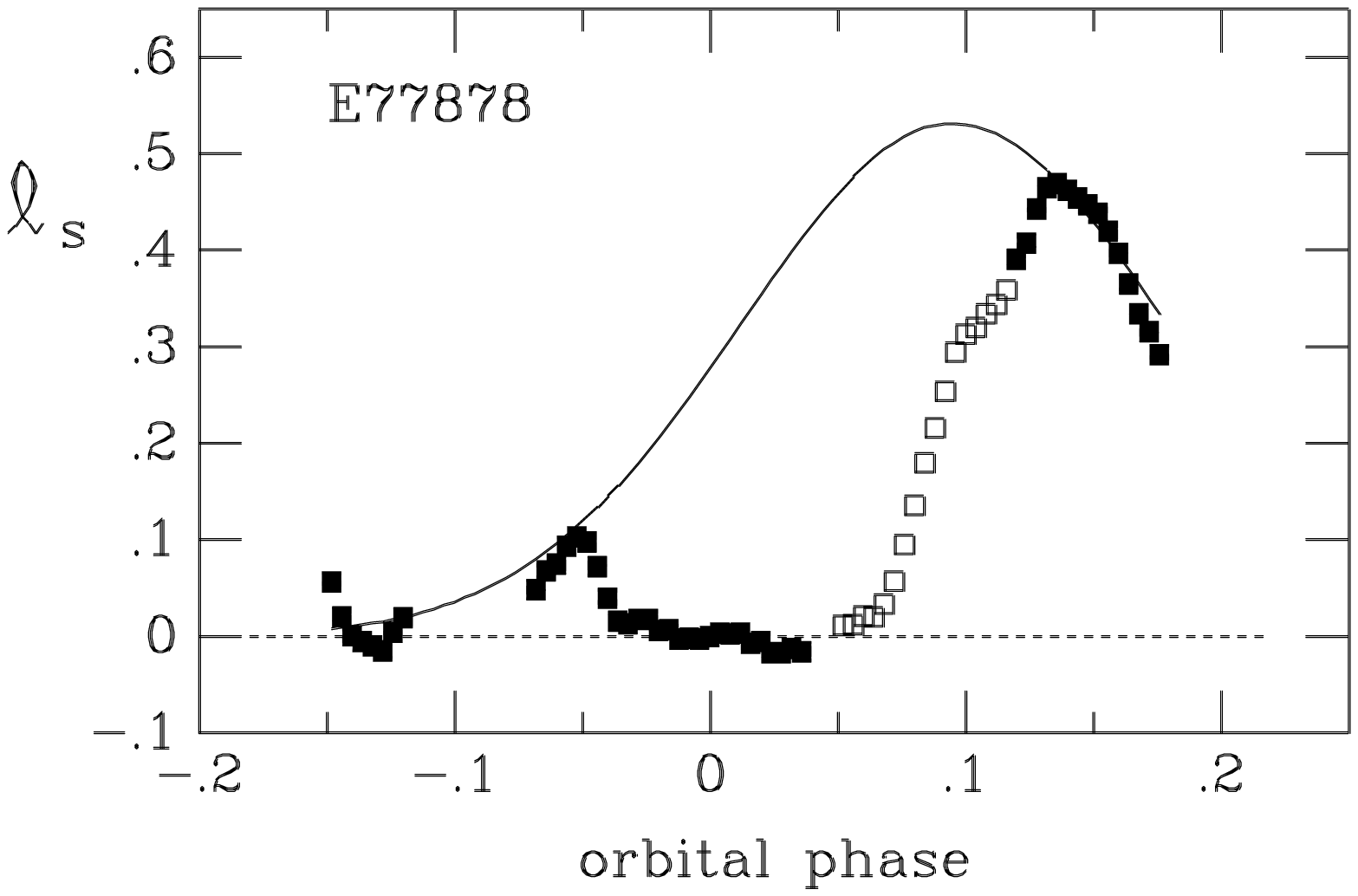} 
\vskip 5truemm
\FigCap { The superhump eclipse light curve of E77878. See text for details. }
\end{figure}

\begin{figure}[htb]
\epsfysize=4.0cm 
\hspace{4.5cm}
\epsfbox{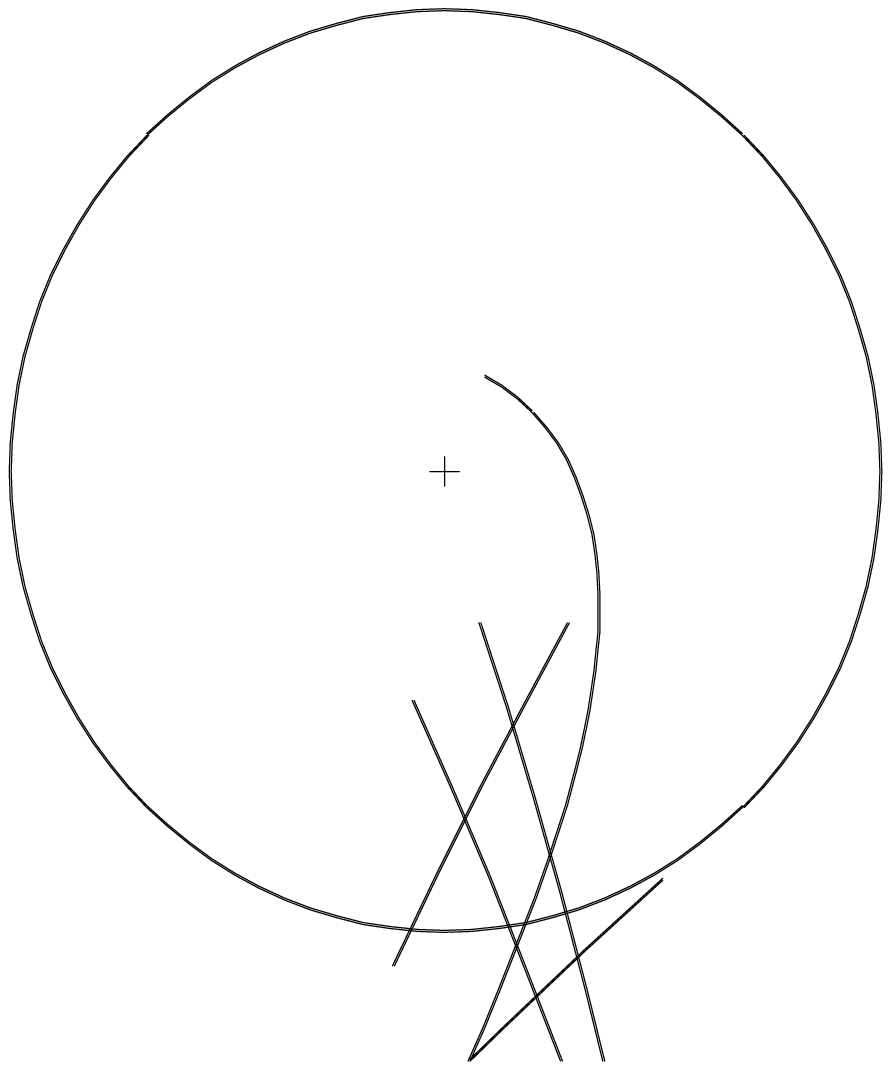} 
\vskip 5truemm
\FigCap { Geometrical constraints on the location of SLS from  
eclipse E77878. Shown are: the disk, the stream trajectory, 
and short sections of the limb of the secondary (in projection on 
the orbital plane) at the four phases of contacts. } 
\end{figure}

We now use the four phases of contacts: $\phi =-0.05, -0.03, 0.06$, 
and 0.12, to determine the location of SLS. For this purpose we employ 
the standard method used commonly (e.g. Wood et al. 1989) for hot spots. 
The result is shown in Fig.8. As we can see, the area defined by the 
four arches, representing the limb of the secondary in projection on 
the orbital plane at those four phases, coincides with the standard 
location of the hot spot at the intersection of the stream trajectory 
with the outer edge of the disk. This, incidentally, is consistent 
with earlier results of Warner and O'Donoghue discussed 
in the Introduction. 

At this point we can explain the absence of the first minimum in the 
observed curve of E77878.  
This eclipse differs from the remaining ones in only one respect: 
it took place well before superhump maximum, i.e. at the time when absorption 
effects, due to the stream overflow, were -- evidently -- not yet present. 
This confirms the existence of an intrinsic connection between superhumps and 
the substantial stream overflow.

\section { Conclusions }

Evidence presented above leads to the following conclusions: 

{\parskip=0truept {

(1) Superhumps are due to periodically enhanced dissipation of the kinetic energy 
of the stream resulting from strongly modulated mass transfer rate. 

(2) Substantial stream overflow occurs around superhump maximum making the superhump 
light source similar to "peculiar" spots observed at intermediate beat phases 
(cf. Smak 2007). 

}}
\parskip=12truept

\begin {references}

\refitem {Hessman, F.V.} {1999} {\ApJ} {510} {867} 

\refitem {Kunze, S., Speith, R., Hessman, F.V.} {2001} {\MNRAS} {322} {499} 

\refitem {O'Donoghue, D.} {1990} {\MNRAS} {246} {29}

\refitem {Smak, J.} {1994} {\Acta} {44} {45}

\refitem {Smak, J.} {2007} {\Acta} {57} {87}

\refitem {Smak, J.} {2008} {\Acta} {58} {55}

\refitem {Warner, B., O'Donoghue, D.} {1988} {\MNRAS} {233} {705} 

\refitem {Wood, J.H., Horne, K., Berriman, G., Wade, R.A.} {1989} {\ApJ} {341} {974} 

\end {references}
\end{document}